\documentclass[onecolumn,showpacs,11pt]{revtex4}
\usepackage{graphicx}
\usepackage{dcolumn}
\usepackage{bm}
\usepackage{epstopdf}
\usepackage{multirow}
\usepackage{eurosym}
\usepackage{wrapfig}
\usepackage{amsfonts}
\usepackage{amsmath}
\usepackage{hyperref}
\usepackage{graphicx,color}
\usepackage{amssymb}
\usepackage[english]{babel}
\usepackage{graphicx}
\usepackage{epsfig}
\usepackage{longtable}
\usepackage{verbatim}
\usepackage{longtable}
\usepackage{color}
\usepackage{subfigure}
\usepackage[utf8]{inputenc}
\begin{document}
\newcommand{\hs}{\hspace*{0.3cm}}
\newcommand{\vs}{\vspace*{0.3cm}}
\newcommand{\be}{\begin{equation}}
\newcommand{\ee}{\end{equation}}
\newcommand{\bea}{\begin{eqnarray}}
\newcommand{\eea}{\end{eqnarray}}
\newcommand{\ben}{\begin{enumerate}}
\newcommand{\een}{\end{enumerate}}
\newcommand{\bde}{\begin{widetext}}
\newcommand{\ede}{\end{widetext}}
\newcommand{\nn}{\nonumber}
\newcommand{\crn}{\nonumber \\}
\newcommand{\Tr}{\mathrm{Tr}}
\newcommand{\non}{\nonumber}
\newcommand{\noi}{\noindent}
\newcommand{\al}{\alpha}
\newcommand{\la}{\lambda}
\newcommand{\bet}{\beta}
\newcommand{\ga}{\gamma}
\newcommand{\va}{\varphi}
\newcommand{\om}{\omega}
\newcommand{\pa}{\partial}
\newcommand{\+}{\dagger}
\newcommand{\fr}{\frac}
\newcommand{\bc}{\begin{center}}
\newcommand{\ec}{\end{center}}
\newcommand{\Ga}{\Gamma}
\newcommand{\de}{\delta}
\newcommand{\De}{\Delta}
\newcommand{\ep}{\epsilon}
\newcommand{\varep}{\varepsilon}
\newcommand{\ka}{\kappa}
\newcommand{\La}{\Lambda}
\newcommand{\si}{\sigma}
\newcommand{\Si}{\Sigma}
\newcommand{\ta}{\tau}
\newcommand{\up}{\upsilon}
\newcommand{\Up}{\Upsilon}
\newcommand{\ze}{\zeta}
\newcommand{\ps}{\psi}
\newcommand{\Ps}{\Psi}
\newcommand{\ph}{\phi}
\newcommand{\vph}{\varphi}
\newcommand{\Ph}{\Phi}
\newcommand{\Om}{\Omega}

\title{Lepton masses and mixings in a $T'$ flavoured 3-3-1 model \\
with type I and II seesaw mechanisms}

\author{V. V. Vien$^{a, b}$}
\email{wvienk16@gmail.com}
\author{H. N. Long$^{c}$}
\email{hnlong@iop.vast.ac.vn}
\author{A. E. C\'arcamo Hern\'andez${}^{d}$}
\email{antonio.carcamo@usm.cl}
\affiliation{$^{{a}}$ Institute of Research and Development, Duy Tan University, 182
Nguyen Van Linh, Da Nang City, Vietnam\\
$^{{b}}$ Department of Physics, Tay Nguyen University, 567 Le Duan, Buon Ma
Thuot, DakLak, Vietnam,\\
$^{{c}}$Institute of Physics, Vietnam Academy of Science and Technology, 10
Dao Tan, Ba Dinh, Hanoi Vietnam\\
$^{{d}}$Universidad T\'{e}cnica Federico Santa Mar\'{\i}a and Centro Cient%
\'{\i}fico-Tecnol\'{o}gico de Valpara\'{\i}so, \\
Casilla 110-V, Valpara\'{\i}so, Chile}

\date{\today}

\begin{abstract}
We propose a renormalizable $T'$ flavor model based on the $SU(3)_C\times SU(3)_L\times U(1)_X\times U(1)_{\mathcal{L}}$ gauge symmetry, consistent with the observed pattern of lepton masses and mixings. The small masses of the light active neutrinos are produced
 from an interplay of type I and type II seesaw mechanisms, which are induced by three heavy right-handed Majorana neutrinos
and three $SU(3)_L$ scalar antisextets, respectively. Our model is only viable for the scenario of normal neutrino mass hierarchy, where the obtained physical observables of the lepton sector are highly consistent with the current neutrino oscillation experimental data. In addition,
 our model also predicts an effective Majorana neutrino mass parameter
  of $m_{\beta} \sim  1.41541\times 10^{-2}$ eV, a Jarlskog invariant of the order of $J_{CP}\sim -0.032$ and a leptonic Dirac CP violating phase of $\de = 238^\circ$, which is inside the $1\sigma$ experimentally allowed range.
\end{abstract}


 \maketitle

\section{Introduction}
Despite its striking consistency with experimental data, the Standard Model (SM) of the elementary particle
physics cannot provide a satisfactory explanation of the fermion mass hierarchy and mixing angles. There is a huge gap of about 5 orders of magnitude
 between the electron and the top quark masses. In addition many experiments show that neutrinos have tiny masses, of about 8 orders of magnitude
  much smaller than the electron mass. Furthermore, the absolute neutrino mass scale as well as the sign of  $\De m^2_{31}$ is still unknown.
   In addition quark mixing angles are small, whereas two of the leptonic mixing angles are large and the other is Cabibbo sized.
    In addition, the existence of three fermion families, which is not explained in the context of the SM, can be understood in the framework
of $SU(3)_{C}\otimes SU(3)_{L}\otimes U(1)_{X}$ models (3-3-1 models), where $U(1)_{X}$ nonuniversal family
symmetry distinguishes the third fermion family from the first and
second ones \cite{Georgi:1978bv, Singer:1980sw,Valle:1983dk,Foot:1994ym,Hoang:1995vq,Hoang:1996gi,CarcamoHernandez:2017kra,CarcamoHernandez:2017cwi}. These models are very important because of the following reasons: 1) The existence of
three generations of fermions arise from the cancellation of chiral anomalies
 and the asymptotic freedom in QCD. 2) The non universal $U(1)_{X}$ symmetry allows to explain the large mass
hierarchy between the heaviest quark family and the two lighter ones. 3) These models provide an explanation for the electric charge quantization \cite{deSousaPires:1998jc,VanDong:2005ux}.
4) CP violation is generated in the 3-3-1 models \cite{Montero:1998yw,Montero:2005yb}. 5) The 3-3-1 models predict the upper
 bound $\sin ^{2}\theta _{W}<\fr{1}{4}$ for the Weinberg mixing angle. 6) Third, these models include a natural Peccei-Quinn
symmetry, which is crucial for addressing the strong-CP problem as explained in detail in Refs. \cite%
{Pal:1994ba,Dias:2002gg,Dias:2003zt,Dias:2003iq}. 7) Models with heavy sterile neutrinos include cold dark matter candidates as weakly
interacting massive particles (WIMPs) \cite{Mizukoshi:2010ky}. A concise review of WIMPs in 3-3-1 Electroweak Gauge Models
 is provided in Ref. \cite{daSilva:2014qba}.

The global fits of the available data from neutrino oscillation
experiments Daya Bay \cite{An:2012eh}, T2K \cite{Abe:2011sj}, MINOS \cite%
{Adamson:2011qu}, Double CHOOZ \cite{Abe:2011fz} and RENO \cite{Ahn:2012nd},
set constraints on the allowed values of the neutrino mass squared splittings, the leptonic mixing parameters and the leptonic Dirac CP violating phase,
as displayed in Table \ref{PLB2018Salas}  (based on Ref. \cite{PLB2018Salas}) for the normal (NH) and inverted (IH) hierarchies of the neutrino mass
spectrum. These facts might suggest that the tiny neutrino masses can be
related to a scale of new physics that, in general, is not related to the
scale of Electroweak Symmetry Breaking $v=246$ GeV. Furthermore, the charged fermion masses can be accommodated in the SM,
at the price of having an unnatural tuning among its different Yukawa couplings. All these unexplained issues within the context of the SM, suggest
 that new physics have to be invoked to address the fermion puzzle of the SM.

The unexplained flavor puzzle of the SM has stimulated work on flavor symmetries which includes the $T^{\prime }$ \cite{Tp1,Tp3,Tp4,Tp5,Tp6,Tp7,Tp8, Tp9, Tp10,Tp13,Aranda:2000tm,Sen:2007vx,Aranda:2007dp,Chen:2007afa,Frampton:2008bz,Eby:2011ph,Frampton:2013lva,Chen:2013wba} discrete groups, that are used in order to provide an explanation for the observed pattern of SM fermion masses and mixing angles.
In this paper we propose a 3-3-1 model with neutral leptons based on $T'$ flavor symmetry consistent with the current neutrino oscillation experimental data of
Ref. \cite{PLB2018Salas} for the scenario of normal hierarchy. The masses of the light active neutrinos are generated from an interplay of type I and type II seesaw mechanisms mediated by three heavy right-handed Majorana neutrinos and three $SU(3)_L$ scalar antisextets, respectively.

Despite the $T'$ has been previously studied in Refs. \cite%
{Tp1,Tp3,Tp4,Tp5,Tp6,Tp8, Tp9, Tp10,Tp13,Aranda:2000tm,Sen:2007vx,Aranda:2007dp,Chen:2007afa,Frampton:2008bz,Eby:2011ph,Frampton:2013lva,Chen:2013wba}, to the best of our knowledge, this discrete group has not been considered before in this kind of 3-3-1 model.
\begin{table}[h]
\begin{tabular}{|c|c|c|c|c|c|c|c|}
\hline \multicolumn{2}{|c|}
{Parameter}& $\De m_{21}^{2}$($10^{-5}$eV$^2$) & $\De m_{31}^{2}$($10^{-3}$eV$^2$) & $\sin ^{2}\theta _{12}$ & $\sin ^{2}\theta _{23}$ & $\sin ^{2}\theta
_{13}$&$\delta/\pi$ \\ \hline \hline
\multirow{2}{2.3cm}{\hfill Best fit $\pm$ $1\sigma$ \hfill } & NH &$7.55^{+0.20}_{-0.16}$ & $2.50\pm 0.03$ & $0.320^{+0.020}_{-0.016}$ & $0.547^{+0.020}_{-0.030}$ & $0.0216^{+0.00083}_{-0.00069}$&$1.32^{+0.21}_{-0.15}$ \\
\cline{2-8}   & IH& $7.55^{+0.20}_{-0.16}$ & $-2.42^{+0.03}_{-0.04}$  & $0.320^{+0.020}_{-0.016}$ &  $0.551^{+0.018}_{-0.030}$  & $0.0222^{+0.00074}_{-0.00076}$ &$1.56^{+0.13}_{-0.15}$\\  \hline
\end{tabular}
\caption{The experimental best fit values of neutrino mass squared splittings
and leptonic mixing parameters, taken from Ref. \cite{PLB2018Salas}. Here, NH and IH stand for Normal Hierarchy and Inverted Hierarchy, respectively.}
\label{PLB2018Salas}
\end{table}\\
as follows. In Sec. \ref{model} we describe our $T'$ flavor 3-3-1 model, which contains several scalar fields introduced to explain the lepton masses and mixings. The results of our numerical analysis
are presented in Sec. \ref{result}. Our conclusions are stated in section \ref{conclus}. Appendix \ref{Tpbreaking3} provides the breakings of $T'$ by a scalar field in the triplet representation $\underline{3}$ of this discrete group.

\section{\label{model} The model}
We consider a model based on the $SU(3)_C\times SU(3)_L\times U(1)_X\times U(1)_{\mathcal{L}}$ gauge symmetry, which is supplemented by the $T'$ discrete group, introduced to generate a viable pattern of lepton masses and mixings consistent with the current neutrino oscillation experimental data. The lepton assignments of the model, under the $[\mathrm{SU}(3)_L,
\mathrm{U}(1)_X, \mathrm{U}(1)_\mathcal{L},\underline{T}']$ symmetries, are given in Tab. \ref{Lepcont},
\begin{table}[ht]
\begin{center}
\begin{tabular}{|c||c|c|c|c|c|c|c|c|c|c|c|c|}
\hline
Fields & $\psi_{1 (\al) L}$ &$l_{1 (\al)R}$&$\phi$&$\phi'$&$\si$&$\si'$&$s$&$\rho$\\ \hline\hline
$\mathrm{SU}(3)_L$  & $3$ &$1$&$3$&$3$&$6^*$&$6^*$&$6^*$&$3$   \\\hline
$\mathrm{U}(1)_X$  & $-\fr{1}{3}$ &$-1$&$\fr{2}{3}$&$\fr{2}{3}$ &$\fr{2}{3}$&$\fr{2}{3}$&$\fr{2}{3}$ &$\fr{2}{3}$\\ \hline
$\mathrm{U}(1)_\mathcal{L}$ &  $\fr{2}{3}$ &$1$&$-\fr{1}{3}$&$-\fr{1}{3}$&$-\fr{4}{3}$&$-\fr{4}{3}$&$-\fr{4}{3}$&$-\fr{4}{3}$  \\\hline
$\underline{T}'$&  $\underline{1} (\underline{2})$  &$\underline{1} (\underline{2})$&$\underline{1} $&$\underline{3}$&$\underline{1}$&$\underline{2}$&$\underline{3}$&$\underline{1}$  \\ \hline
\end{tabular}
\caption{\label{Lepcont} The lepton and scalar assignments under $SU(3)_L\times U(1)_X\times U(1)_{\mathcal{L}}\times T'$.}
\end{center}
\end{table}
where $\al=2,3$ is a family
index of the last two lepton generations,
which defines the components of the $T'$ doublet representations.

To generate masses for the charged leptons, we need two $SU(3)_L$ scalar
multiplets, namely $\phi$ and $\phi'$, whose assignments under the different discrete group factors of the model are given in Table. \ref{Lepcont}.
With the particle content and symmetries specified in Table. \ref{Lepcont}, the following Yukawa interactions for charged leptons arise:
\be -\mathcal{L}_{l}=h_{1}\bar{\psi}_{1 L} \phi l_{1R}+
h_2(\bar{\psi}_{\al L} \phi)_{2} l_{\al R}+h_{3}(\bar{\psi}_{\al L} \phi')_{2} l_{\al R}+H.c. \label{Llep}\ee
In this work, we impose the $T'\rightarrow Z_2$ symmetry breaking chain, which gives rise to the VEV pattern $\langle \phi' \rangle=(\langle \phi'_1 \rangle, 0, 0$) for the $T'$ triplet scalar $\phi'$. In addition, the VEVs of the $SU(3)_L$ scalars $\phi'_1$ and $\phi$ are given by:
\bea
 \langle \phi'_1 \rangle &=& \left(  0 \hs
  v' \hs
  0\right)^T, \hs \langle \phi \rangle = \left( 0 \hs
  v \hs
  0 \right)^T. \label{vevphiphip} \eea
Then, after electroweak symmetry breaking, the following charged lepton mass terms are obtained:
\bea
-\mathcal{L}^{\mathrm{mass}}_l&=&h_{1}v_1\bar{l}_{1 L} l_{1R}
-(h_{2}v-h_3v')\bar{l}_{2L} l_{2 R}
+(h_{2}v+h_3v')\bar{l}_{3L}l_{3 R}+H.c, \eea
From the mass terms given above, we find that the SM charged lepton mass matrix is diagonal and the masses for the SM charged leptons are given by:
\bea
 m_e&=& h_1v, \hs m_\mu= -h_{2}v+h_3v', \hs m_\tau= h_{2}v+h_3v',\label{Mlep}\eea
and thus the diagonalization
matrices are $U_{lL}= U_{lR}=1$. This means that the
charged lepton fields $l_{1,2,3}$ by themselves are physical mass
eigenstates, and the Pontecorvo–Maki–Nakagawa–Sakata (PMNS) leptonic mixing matrix is the rotation matrix that diagonalizes the light active neutrino mass matrix. The masses of
muon and tau leptons are explicitly separated by $\phi'$ resulting from
the breaking $T'\rightarrow Z_6$. This is why we introduce
$\phi'$ in accompanying with $\phi$.
For the charged leptons masses at the electroweak scale we use the values given in Particle Data Group 2018 \cite{PDG2018}: $ m_e \simeq 0.511\ \textrm{MeV},\,  m_{\mu}\simeq 105.66 \ \textrm{MeV},\, m_{\tau}\simeq1776.86\ \textrm{MeV}$.
 Thus, we get \[ h_1 v=0.511\ \textrm{MeV},\ h_2v =835.6\
\textrm{MeV},\   h_3 v'=941.26 \ \textrm{MeV} . \]
Consequently, to explain the SM charged lepton mass hierarchy, it is required that $|h_1|\ll |h_2|$, and if $|v'|\sim |v|$, then $|h_2|$ and $|h_3|$ have to be of the same order of magnitude.

To generate the masses of the light active neutrinos we introduce six $SU(3)_L$ scalar antisextets, namely, $\si , \si'_k (k=1,2), s_j (j=1,2,3)$ and
one $SU(3)_L$ scalar triplet $\rho$. The $SU(3)_L$ scalar antisextet $\si$ is assigned as a $T'$ trivial singlet,
 whereas the $\si'_k (k=1,2), s_j (j=1,2,3)$ scalar fields are grouped into a $T'$ doublet and a $T'$ triplet, respectively. Furthermore, the $SU(3)_L$ scalar
  triplet $\rho$ is assigned as a trivial $T'$ singlet. The lepton and scalar field assignments under the different group factors
   of the model are shown in Tab.\ref{Lepcont}.
In this work we assume that both $T'\rightarrow Z_6$ and
$Z_6\rightarrow \{\mathrm{identity}\}$  breakings must take place in the neutrino sector. The breakings $T'\rightarrow Z_6$ can be achieved by the scalar antisextet $s$
whose VEV pattern is set as $(\langle s\rangle, 0, 0)$ under $T'$, where
 \be
\langle s\rangle =\left(%
\begin{array}{ccc}
  \la_{s} & 0 & v_{s} \\
  0 & 0 & 0 \\
  v_{s} & 0 & \Lambda_{s} \\
\end{array}%
\right). \label{vevs}\ee
To achieve the direction of the $Z_6\rightarrow
\{\mathrm{identity}\}$ breaking chain, we additionally introduce another scalar assigned as $\underline{2}$ under $T'$. We can therefore
  understand the misalignment of the
VEVs as follows. The $T'$ discrete group is spontaneously broken via two stages, the first
stage is $T'\rightarrow Z_6$ and the second one is
$Z_6\rightarrow \{\mathrm{identity}\}$. The second stage can be achieved
by adding a new $\mathrm{SU}(3)_L$ anti-sextet $\si'$, transforming as
$\underline{2}$ under $T'$ as shown in Table \ref{Lepcont}, with
VEVs chosen as
\[
\langle \si'_1\rangle=\langle \si'_2\rangle=\left(%
\begin{array}{ccc}
  \la'_{\si } & 0 & v'_{\si} \\
  0 & 0 & 0 \\
  v'_{\si} & 0 & \Lambda'_{\si} \\
\end{array}%
\right).
\]
On the other hand, the neutrino
Yukawa interactions invariant under the symmetries of our model are given by:
\bea -\mathcal{L}_\nu&=& \fr 1 2 x \bar{\psi}^c_{1 L} \sigma\psi_{1 L}
+ y \left(\bar{\psi}^c_{1 L} \sigma'\right)_{\underline{2}}\psi_{\al L}
+\fr 1 2 z (\bar{\psi}^c_{\al L}s)_{\underline{2}} \psi_{\al L}
+\fr 1 2 t \left(\bar{\psi}^c_{\al L} \rho\right)_{\underline{2}}\psi_{\al L}
+H.c.\label{Llep}\eea
After electroweak symmetry breaking, we find the following neutrino mass terms:
\bea
 -\mathcal{L}_\nu^{mass} &=& \fr 1 2 x\la_{\si} \bar{\nu}^c_{1L} \nu_{1L}+\fr 1 2 xv_{\si}\bar{\nu}^c_{1L}N^c_{1R}
 + \fr 1 2 x v_{\si}  \bar{N}_{1R}\nu_{1L} + \fr 1 2 x\Lambda_{\si} \bar{N}_{1R}N^c_{1R}\crn
&+& y \la'_{\si}\bar{\nu}^c_{1L} \nu_{3L}+ y  v'_{\si}\bar{\nu}^c_{1L}N^c_{3R}+  y v'_{\si} \bar{N}_{1R}\nu_{3L}
 + y \Lambda'_{\si}\bar{N}_{1R}N^c_{3R} \crn
&-& y \la'_{\si} \bar{\nu}^c_{1L}\nu_{2L}- y v'_{\si} \bar{\nu}^c_{1L}N^c_{2R}- y v'_{\si} \bar{N}_{1R}\nu_{2L}-
 y\Lambda'_{\si} \bar{N}_{1R}N^c_{2R}\crn
&+& \fr 1 2 z \la_{s}\bar{\nu}^c_{2L}\nu_{3L}+\fr 1 2 z v_{s}\bar{\nu}^c_{2L}N^c_{3R}
+\fr 1 2 z v_{s}\bar{N}_{2R}\nu_{3L}+ \fr 1 2 z \Lambda_{s}\bar{N}_{2R}N^c_{3R}\crn
&+&\fr 1 2 z \la_{s} \bar{\nu}^c_{3L}\nu_{2L}+ \fr 1 2 z v_{s}\bar{\nu}^c_{3L}N^c_{2R}
 + \fr 1 2 z  v_{s}  \bar{N}_{3R}\nu_{2L}+ \fr 1 2 z  \Lambda_{s} \bar{N}_{3R}N^c_{2R}\crn
&+& \fr 1 2 t v_\rho\bar{\nu}^c_{2 L}N^c_{3R}-\fr 1 2 t v_\rho\bar{\nu}^c_{3 L}N^c_{2R}-
\fr 1 2 t v_\rho\bar{N}_{2 R}\nu_{3L}+\fr 1 2 t v_\rho \bar{N}_{3 R}\nu_{2L}+H.c.
\label{Llepmass}\eea
The neutrino mass term $-\mathcal{L}^{\mathrm{mass}}_\nu$  in Eq. (\ref{Llepmass}) can be rewritten in a matrix form as follows:
 \be -\mathcal{L}^{\mathrm{mass}}_\nu=\fr 1 2
\bar{\chi}^c_L M_\nu \chi_L+ H.c.,\label{nm}\ee where
\bea
\chi_L&=&
\left(%
\begin{array}{c}
  \nu_L \\
  N^c_R \\
\end{array}%
\right),\hs\hs\,\,\,\,\, M_\nu =\left(%
\begin{array}{cc}
  M_L & M^T_D \\
  M_D & M_R \\
\end{array}%
\right), \label{mLDR}\\
\nu_L &=& (\nu_{1L},\nu_{2L},\nu_{3L})^T, \,\, N_R=(N_{1R},N_{2R},N_{3R})^T,
\eea
and
\be M_{L,R, D}=\left(%
\begin{array}{ccc}
a_{L,D,R} & -b_{L,D, R} &\hs b_{L,D, R} \\
  -b_{L,D, R} & 0 & c_{L,D, R}+d_{L,D, R} \\
   b_{L,D, R} & c_{L,D, R}-d_{L,D, R} & 0 \\
\end{array}%
\right),\label{Mnu}\ee
with
\bea
a_{L}&=&\la_{\si} x, \,\, b_{L}=y \la'_{\si},\,\, c_{L}=z \la_{s}, \,\, d_{L}=0,\crn
a_{D}&=&v_{\si} x, \,\, b_{D}=y v'_{\si}, \,\, c_{D}=z v_{s}, \,\, d_{D}= t v_\rho,\crn
a_{R}&=&\La_{\si} x, \,\, b_{R}=y \La'_{\si}, \,\, c_{R}=z \La_{s}, \,\, d_{R}=0.
\label{abcdLDR}
\eea
Three light active neutrinos gain masses from a combination of type I and type II seesaw mechanisms as follows from
 Eqs. (\ref{mLDR}) and (\ref{Mnu}). Then, the light active neutrino mass matrix takes the form:
\be
M_{\mathrm{eff}}=M_L-M_D^TM_R^{-1}M_D=M^0_{\mathrm{eff}} +dM_1+dM_2, \label{Meff}\ee
where
\bea
M^0_{\mathrm{eff}}&=&\left(%
\begin{array}{ccc}
 A  & -B &B \\
 -B & D+H & D \\
 B  & D & D+H \\
\end{array}%
\right),\hs  dM_1=\left(%
\begin{array}{ccc}
 0  & p &p \\
 p & q & 0 \\
 p  & 0 & -q \\
\end{array}%
\right), \hs dM_2=\left(%
\begin{array}{ccc}
 0  & 0 &0 \\
 0 & s & r \\
 0  & r & s \\
\end{array}%
\right), \label{dm1dm2}\eea
 with
\bea
 A&=&\la_{si} x+\fr{x \left[2 v'_{\si} (\La_{si} v'_{\si} - 2 \La'_{\si} v_{\si}) y^2 - \La_s v_{\si}^2 x z\right]}{2 \La_{\si}^{'2} y^2 + \La_s \La_{\si} x z}, \crn
B&=&\la'_{\si} y-\fr{2 \La'_{\si} v_{\si}^{'2} y^3 + (\La_{\si} v'_{\si} v_s + \La_s v'_{\si} v_{\si} - \La'_{\si} v_s v_{\si}) x y z}{2 \La_{\si}^{'2} y^2 + \La_s \La_{\si} x z},\crn
 D&=&\la_s z -\fr{z \left[(\La_s^2 v_{\si}^{'2} - 2 \La'_{\si} \La_s v'_{\si} v_s - \La_{\si}^{'2} v_s^2) y^2 - \La_s \La_{\si} v_s^2 x z\right]}{\La_s(2 \La_{\si}^{'2} y^2 + \La_s \La_{\si} x z)},\crn
H&=&- \la_s z+\fr{z \left[-2 v'_{\si} (\La_s v'_{\si} - 2 \La'_{\si} v_s) y^2 + \La_{\si} v_s^2 x z\right]}{2 \La_{\si}^{'2} y^2 + \La_s \La_{\si} x z}, \label{ABDH}\\
p&=&\fr{t v_{\rho} (-\La_{\si} v'_{\si} + \La'_{\si} v_{\si}) x y}{2 \La_{\si}^{'2} y^2 + \La_s \La_{\si} x z}, \hs \,\, q =\fr{2 \La'_{\si} t v_{\rho} (-\La_s v'_{\si} + \La'_{\si} v_s) y^2}{\La_s(2 \La_{\si}^{'2} y^2 + \La_s \La_{\si} x z)}, \label{pq}\\
s&=&-\fr{\La_{\si}^{'2} t^2 v_{\rho}^2 y^2}{2 \La_{si}^{'2} \La_s y^2 z + \La_s^2 \La_{\si} x z^2},\,\,\,\,\, r=\fr{t^2 v_{\rho}^2 (\La_{\si}^{'2} y^2 + \La_s \La_{\si} x z)}{\La_s z (2 \La_{\si}^{'2}y^2 + \La_s \La_{\si} x z)}  \label{rs}
\eea
Let us note that, as indicated by Eq. (\ref{Meff}), the light active neutrino mass matrix $M_{\mathrm{eff}}$ receives a contribution from the three $SU(3)_L$ scalar antisextets, i.e., $\si, \si'$ and $s$, namely $M^0_{\mathrm{eff}}$ as well as contributions $dM_1$ and $dM_2$ arising from the $SU(3)_L$ scalat triplet $\rho$. In the case where the $\rho$ contribution is
forbidden, the two matrices  $dM_1$ and $dM_2$ will vanish, and hence the matrix
$M_{\mathrm{eff}}$ in Eq. (\ref{Meff}) reduces to $M^0_{\mathrm{eff}}$. As will shown below,  $M^0_{\mathrm{eff}}$ can
approximately fit the data with $\theta_{13}=0$ that can be considered as a leading order approximation for
the recent neutrino experimental data. The second and the third terms, which correspond to the contributions of the $\rho$ triplet
will generate the Cabibbo sized deviation from $\theta_{13}=0$, thus giving rise to the
experimental value of the reactor mixing angle $\theta_{13}$. Thus, in this work we consider the $\rho$ contribution as a small
perturbation ($d_D\ll d_R$) needed to generate the Cabibbo sized value of the reactor mixing angle measured by the neutrino oscillation experiments. On the other hand, since $p,q$ are proportional to $d_D$ whereas $s, r$ are proportional to $d^2_D$,
 we can work in the limit $r,s \ll 1$, and safely neglect the second order correction $dM_2$ to the light active neutrino mass matrix.

The first term in Eq. (\ref{Meff}) has three exact eigenvalues given by
\be m_{1,2}
=\fr 1 2 \left(A+H \mp \sqrt{(A-H)^2+8 B^2}\right),\hs
m_3 =2 D+H,\label{la123}\ee and the corresponding eigenstates included in the
lepton mixing matrix take the form: \be U_0= \left(%
\begin{array}{ccc}
  \fr{K}{\sqrt{K^2+2}} & \fr{\sqrt{2}}{\sqrt{K^2+2}} & 0 \\
  -\fr{1}{\sqrt{K^2+2}} &\fr{1}{\sqrt{2}}\fr{K}{\sqrt{K^2+2}} & \fr{1}{\sqrt{2}} \\
  \fr{1}{\sqrt{K^2+2}} & -\fr{1}{\sqrt{2}}\fr{K}{\sqrt{K^2+2}}& \fr{1}{\sqrt{2}} \\
\end{array}%
\right), \hs K=\fr{A-H-\sqrt{(A-H)^2+8B^2}}{2B}.\label{U0}\ee
At the first order of perturbation theory, the matrix $dM_1$ in Eq.
(\ref{Meff}) does not contribute to the eigenvalues of the matrix $M_{\mathrm{eff}}$, however, it changes the corresponding eigenvectors.
 Indeed, the three eigenvalues of the light active neutrino mass matrix $M_{\mathrm{eff}}$ are obtained as follows:
\be
m'_1=m_1,\hs m'_2=m_2,\hs m'_3=m_3, \label{la123p}
\ee
where $m_{1,2,3}$ are given by Eq. (\ref{la123}).
The corresponding perturbed leptonic mixing matrix takes the form:
\bea
U&=&U_0+\De U= U_{0}+\left( \begin{array}{ccc}
 0&\hs 0&\hs\De U_{13} \\
 \De U_{21}&\hs \De U_{22}&\hs\De U_{23} \\
 \De U_{31}&\hs \De U_{32}&\hs \De U_{33} \\
 \end{array}\right),   \label{Ulep}
\eea
where $U_0$ is defined by Eq. (\ref{U0}), and the $\De U_{ij}$ ($i,j=1,2,3$) matrix elements are given by
  \bea
   \De U_{21}&=&\De U_{31}= \fr{K p - q}{\sqrt{K^2  +2}(m_1 - m_3)},\hs
  \De U_{22}=\De U_{32}=\fr{2 p + K q}{\sqrt{2}\sqrt{K^2  +2}(m_2 - m_3)},\crn
  \De U_{13}&=&-\fr{\sqrt{2}\{ [2 m_1 + K^2 (m_2 - m_3) - 2 m_3] p + K (m_1 - m_2) q\}}{(K^2+2) (m_1 - m_3) (m_2 - m_3)},\crn
  \De U_{23}&=&- \fr{2 K (m_1 - m_2) p + K^2 (m_1 - m_3) q + 2 (m_2 - m_3) q}{(K^2+2) (m_1 - m_3) (m_2 - m_3)},\crn
  \De U_{33}&=&\fr{ 2 K (m_1 - m_2) p + K^2 (m_1 - m_3) q + 2 (m_2 - m_3) q}{(K^2+2) (m_1 - m_3) (m_2 - m_3)},\label{dUelement}\eea
  with $p,q$, $m_i = \la_i \, (i=1,2,3)$ and $K$ are given in Eqs. (\ref{pq}), (\ref{la123}) and (\ref{U0}), respectively.

\section{\label{result} Numerical results}

The matrix $U_0$ given by Eq. (\ref{U0}) can be parameterized in terms of three Euler's
angles, satisfying the relations $\theta_{13}=0,\,
\theta_{23}=\pi/4$ and $\tan\theta_{12}=\sqrt{2}/K\equiv  \sqrt{2}B/(m_1-H)$.  In the case $A-H=B<0$, with $B$ being a real number, $\theta_{12}=\fr{\pi}{4}$ and $U_0$ becomes an exact Tri-bimaximal mixing matrix $U_{HPS}$ which can be considered as a zero order approximation for the Pontecorvo–Maki–Nakagawa–Sakata (PMNS) leptonic mixing matrix constrained by the recent
neutrino oscillation experimental data. The recent data imply that $\theta_{13}\neq 0$, however, the contribution arising from $dM_1$ will generate the experimentally observed deviation from $\theta_{13}=0$, thus giving rise to the measured value of the reactor mixing angle.
It is easy to show that our model is consistent with the current neutrino oscillation experimental data
since the experimental values of the six physical observables of the neutrino sector, namely,
the leptonic Dirac CP violating phase, the leptonic mixing angles and the neutrino
mass squared splittings can successfully be reproduced for appropriate values of the neutrino sector model parameters as shown below. Indeed, in the standard parametrization of the PMNS leptonic mixing matrix, the three leptonic mixing
 angles $\theta_{12}, \theta_{23}$ and $\theta_{13}$ can be defined in terms of the elements of the
leptonic mixing matrix as follows:
\bea
t_{12}= U_{12}/U_{11},\,\, t_{23} =U_{23}/U_{33},\,\, s_{13} e^{-i \delta}=U_{13}
\label{SP}
\eea
Then, using from Eqs. (\ref{U0}), (\ref{Ulep}), (\ref{dUelement}) and \ref{SP} we get:
\bea
p&=& \sqrt{2} D s_{13}e^{-i\delta}+B s_{13}e^{-i\delta}\left(t_{12}-\fr{1}{t_{12}}\right)+B\fr{(t_{23}-1)}{t_{23}+1},\crn
q&=&\sqrt{2} B s_{13} e^{-i\delta} +2 D \fr{ (t_{23}-1)}{t_{23}+1 },\hs
H= A +\sqrt{2} B \left(t_{12}-\fr{1}{t_{12}}\right). \label{pqH}
\eea
We found that the inverted hierarchy scenario of our model cannot accommodate the experimental data on neutrino oscillations, however,
 the model predictions in the lepton sector are in
good agreement with the recent neutrino oscillation experimental data for the case of normal hierarchy, which favors the normal hierarchy over the inverted one at $3.4\sigma$. Indeed, for the normal neutrino mass spectrum, taking the best fit values of the leptonic mixing angles and Dirac CP violating phase as well as the neutrino mass-squared differences given in Ref. \cite{PLB2018Salas} as displayed in Tab. \ref{PLB2018Salas},  $s^2_{12}= 0.320 ,\, s^2_{23}= 0.540, \, s^2_{13}= 0.0216$, $\de = 238^o$ and $\De m^2_{21}=m'^2_{2}-m'^2_{1}=7.55\times 10^{-5}\, \mathrm{eV}^2$, $\De m^2_{31}= m'^2_{3} -m'^2_{1}=2.50\times 10^{-3}\ \mathrm{eV}^2$, we find the following solution:
\bea
 A&=&0.545705 B - \fr{0.0000124518}{B},\crn
D&=&0.272853 B+\fr{6.22589\times 10^{-6}}{ B} \crn
  &+&7.39148\times 10^{-6} \sqrt{1.1267\times 10^7 +\fr{0.70948}{B^2}+1.05145\times 10^{10} B^2}.\label{AD}
\eea

\begin{figure}[ht]
\bc
\includegraphics[width=12.5cm, height=18.5cm]{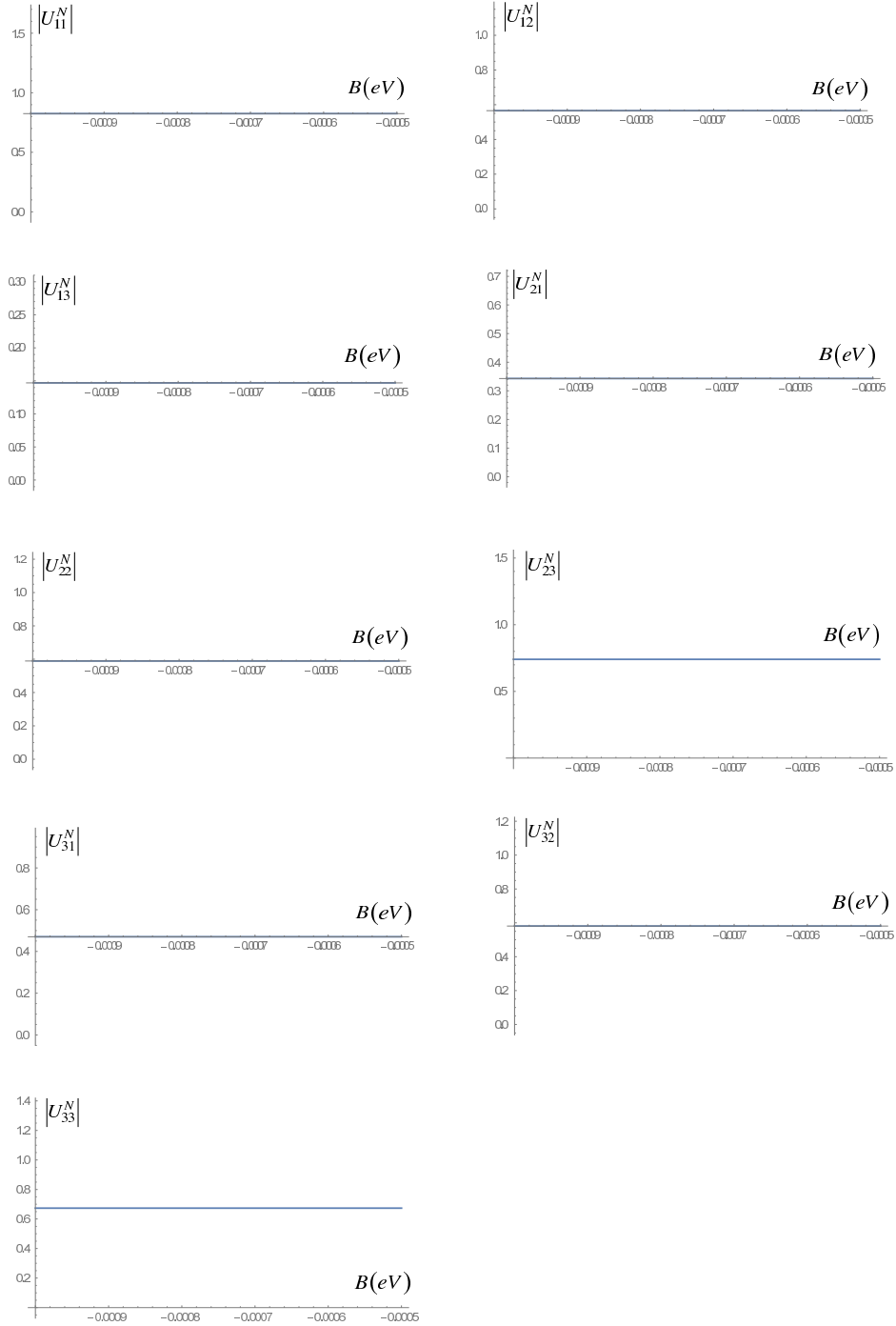}
\vspace*{-0.1cm} \caption[$U^N_{ij}\hspace{0.15 cm} (i,j=1,2,3)$  as functions of $B$  with $B\in (- 10^{-3}, -5\times 10^{-4})$.]{$U^N_{ij}\hspace{0.15 cm} (i,j=1,2,3)$  as functions of $B$  with $B\in (- 10^{-3}, -5\times 10^{-4})$.}\label{UijN}
\ec
\end{figure}

In the scenario of normal hierarchy, the range of the elements $U_{ij} \hspace{0.10 cm} (i,j=1,2,3)$ in Eq. (\ref{Ulep}) are displayed in Fig. \ref{UijN} with $B\in (- 10^{-3}, -5\times 10^{-4}) \, \mathrm{eV}$.

\begin{figure}[ht]
\begin{center}
\includegraphics[width=7.0cm, height=6.5cm]{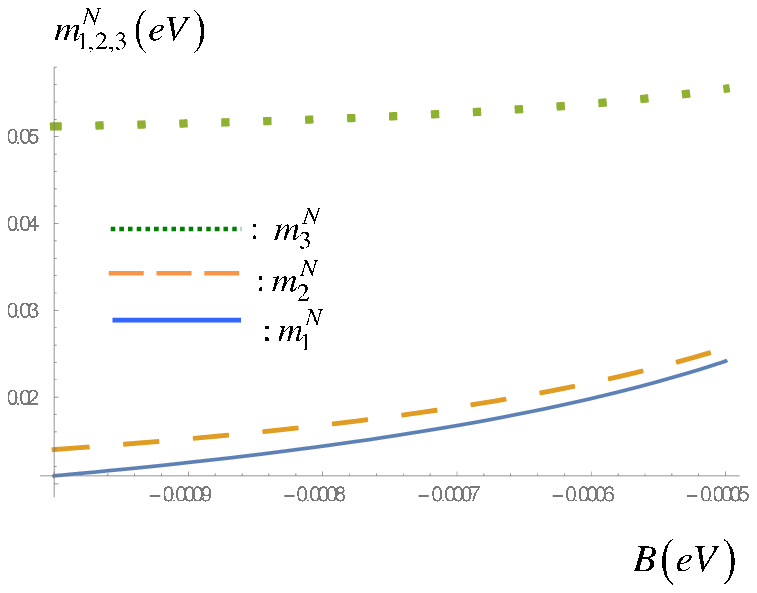}
\vspace*{-0.1cm} \caption[$m_{1,2,3}$ as functions of $B$ in the Normal spectrum with
 $B \in (-5\times 10^{-4}, -10^{-3})\, \mathrm{eV}$.]{$m_{1,2,3}$ as functions of $B$ in the Normal spectrum with
 $B \in (-5\times 10^{-4}, -10^{-3})\, \mathrm{eV}$.}\label{m123N}
\end{center}
\end{figure}

The values of the light active neutrino masses $m_{1,2,3}$ as functions of the effective $B$ parameter with $B \in (- 10^{-3}, -5\times 10^{-4})\, \mathrm{eV}$ are plotted in Fig.\ref{m123N}, for the scenario of normal neutrino mass hierarchy.

\begin{figure}[h]
\begin{center}
\includegraphics[width=7.0cm, height=5.5cm]{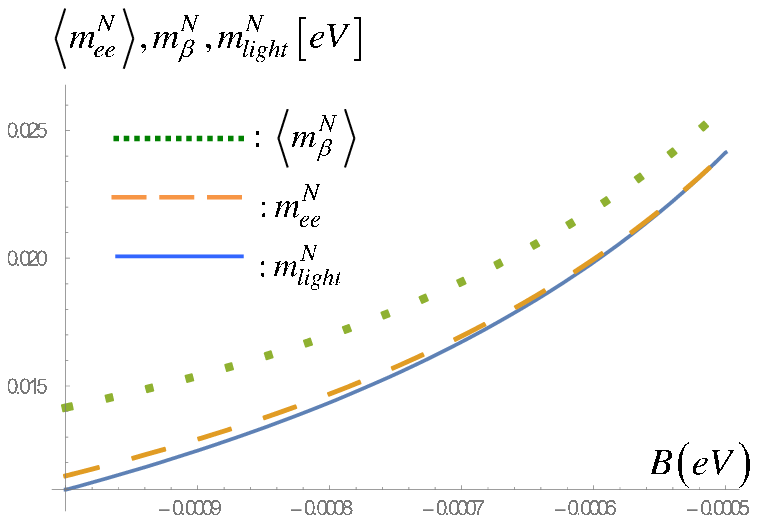}
\vspace*{-0.1cm} \caption[ $\langle m_{ee}\rangle$, $m_{\beta}$ and $m_{light}$ as functions of $B$
with $B \in (-5\times 10^{-4}, -10^{-3})\, \mathrm{eV}$ in the normal spectrum.]{$\langle m_{ee}\rangle$, $m_{\beta}$ and $m_{light}$ as functions of $B$
with $B \in (-5\times 10^{-4}, -10^{-3})\, \mathrm{eV}$ in the normal spectrum.}\label{meeN}
\vspace*{-0.3cm}
\end{center}
\end{figure}

The effective neutrino mass $\langle m_{ee}\rangle$ governing neutrinoless double beta decay
 \cite{betdecay3,betdecay4,betdecay6} takes the form $\langle m_{ee}\rangle=  \left|\sum^3_{i=1} U_{ei}^2 m_i \right|$, whereas $
m_{\beta} = \left\{\sum^3_{i=1} |U_{ei}|^2 m_i^2 \right\}^{1/2}$ where $m_{i}$ and $U_{ei}$ are defined by Eqs. (\ref{U0}),  (\ref{la123p}), (\ref{Ulep}) and (\ref{dUelement}). We plot the parameters $\langle m_{ee}\rangle$ and $m_{\beta}$ in Fig.\ref{meeN} with $B \in ( -10^{-3},-5\times 10^{-4})\, \mathrm{eV}$.
In the case $B=-10^{-3}\, \mathrm{eV}$, the physical neutrino masses and the other parameters are explicitly
 given in Table. \ref{para}. The value of the
Jarlskog invariant $J_{CP}$ which determines the magnitude of CP violation in neutrino
oscillations, in the model under consideration, is determined as \cite{PDG2018} $J_{CP} =
\mathrm{ Im}\left[U_{23} (U_{13})^* U_{12} (U_{22})^*\right]=-3.21528\times 10^{-2}$.
\begin{table}[h]
\caption{ \label{para} The model parameters in the case $B=-10^{-3} \, \mathrm{eV}$  in the normal hierarchy}
{\begin{tabular}{@{}cccccccccc@{}} \toprule
& Parameters $[ \mathrm{eV}]$&\hs The derived values $[ \mathrm{eV}]$&\\\hline
& \,\,$A$ &\, $1.19061\times10^{-2}$&\\
 & \,\,$D$ &\, $1.90922\times 10^{-2}$&\\
  & \,\,$H$ &\, $1.29975\times10^{-2}$&\\
  & \,\,$K$ &\, $2.06155$&\\
  &\,\, $p$ &\, $(-2.23417 + 3.44627)\times 10^{-3}$&\\
  &\,\, $q$ &\, $(1.91002 - 0.17549)\times 10^{-3}$&\\
   & $m^N_{light}\equiv m^N_1$ &\, $1.09359\times 10^{-2}$&\\
    & \,\,$m^N_2$ &\, $1.39676\times10^{-2}$&\\
    & \,\,$m^N_3$ &\, $5.1182 \times10^{-2}$&\\
     & \,\,$\sum m^N_{i} $ &\, $7.60855\times10^{-2}$&\\
      & \,\,$\langle m^N_{ee}\rangle$ &\, $1.14354 \times 10^{-2}$&\\
       & \,\,$m^N_{\beta}$ &\, $1.41541  \times10^{-2}$&\\
\botrule
\end{tabular}}
\end{table}

In the 3-3-1 models, the parameters $\la_{s,\si} \sim \fr{v^2_{s,\si}}{\La_{s,\si}},\, \la'_{\si} \sim \fr{v^{'2}_{\si}}{\La'_{\si}}$ and $\la_{s,\si}, \la'_{\si}$ are at the eV scale \cite{331r6}. Hence, in order to have explicit values for the model
parameters, we assume $\La_{s,\si} =v^2_{s,\si},\, \, \La'_{\si} =v^{'2}_{\si}$ and $v_{\si} =v_s,\, v'_{\si} =-v_s, \, \la'_{\si}=\la_{\si}$. By comparing the expressions of the parameters $A, B, D, H, p, q$ with their corresponding values obtained in Tab. \ref{para}, we get:
\bea
x&=& 0.0714657 + 0.286867 i, \hs\hs y = 0.0423168 - 0.0545414  i ,\crn
z&=& -0.0319395 - 0.000219628  i, \,\,\, t = (0.00190228 - 0.000161596 i) \fr{ v_s}{v_{\rho}}, \crn
\la_{\si}&=&-0.999572 - 0.0252618 i,\hs\, \, \la_s =-0.602386 + 0.0110186  i.\label{xyzla}
\eea

\section{\label{conclus}Conclusions}
We have constructed a $T'$ flavor model based on the $\mathrm{SU}(3)_C \otimes \mathrm{SU}(3)_L \otimes \mathrm{U}(1)_X\otimes\mathrm{U}(1)_\mathcal{L}$ gauge symmetry responsible for lepton masses and mixings.
We argue how flavor mixing patterns and mass splitting are obtained with a perturbed $T'$
symmetry. In the model under consideration, the naturally small neutrino masses arise from a combination of type I and type II seesaw
 mechanisms mediated by three heavy right-handed Majorana neutrinos
and three $SU(3)_L$ scalar antisextets, respectively. Our model predicts normal neutrino mass ordering with the inverted
ordering disfavoured by our fit. In addition, we find an effective Majorana neutrino mass parameter of $m_{\beta} \sim  1.41541\times 10^{-2}$ eV,
a Jarlskog invariant $J_{CP}\sim -0.032$ and a leptonic Dirac CP phase $\de = 238^\circ$ for
 the scenario of normal neutrino mass hierarchy.

\section*{Acknowledgments}
This research is funded by the Vietnam  National Foundation for Science and Technology Development (NAFOSTED)
under grant number 103.01-2017.341, and by Fondecyt (Chile), Grants
No.~1170803, CONICYT PIA/Basal FB0821.

\appendix
\section{\label{Tpbreaking3} The breakings of $T'$ by triplet $3$}
Under $T'$ group, for triplets $\underline{3}$ we have the followings alignments:

\begin{itemize}
\item[(1)] The first alignment: $0=\langle \phi'_2\rangle=\langle \phi'_3\rangle
\neq\langle \phi'_1\rangle$ or $0=\langle \phi'_1\rangle=\langle \phi'_2\rangle
\neq\langle \phi'_3\rangle$ or $0=\langle \phi'_1\rangle=\langle \phi'_3
\rangle\neq\langle \phi'_2\rangle$ then $T'$ is broken into $Z_2$.

\item[(2)] The second alignment: $0=\langle \phi'_1\rangle\neq\langle \phi'_2\rangle
=\langle \phi'_3\rangle \neq 0$ or $0=\langle \phi'_2\rangle\neq\langle \phi'_1\rangle
=\langle \phi'_3\rangle \neq 0$ or $0\neq \langle \phi'_1\rangle=\langle \phi'_2\rangle
\neq\langle \phi'_3\rangle=0$ then $T'$ is broken into $Z_2$.

\item[(3)] The third alignment: $0=\langle \phi'_1\rangle\neq\langle \phi'_2\rangle
\neq\langle \phi'_3\rangle \neq 0$ or $0=\langle \phi'_2\rangle\neq\langle \phi'_1
\rangle\neq\langle \phi'_3\rangle \neq 0$ or $0\neq \langle \phi'_1\rangle\neq\langle \phi'_2\rangle\neq\langle \phi'_3\rangle=0$ then $T'$ is broken
into $Z_2$.

\item[(4)] The fourth alignment: $\langle \phi'_1\rangle\neq\langle \phi'_2\rangle
\neq\langle \phi'_3\rangle$ then $T'$ is broken into $Z_2$.

\item[(5)] The fifth alignment: $0\neq\langle \phi'_1\rangle\neq\langle \phi'_2\rangle
=\langle \phi'_3\rangle\neq0$ or $0\neq\langle \phi'_1\rangle=\langle \phi'_3\rangle
\neq\langle \phi'_2\rangle\neq0$ or $0\neq\langle \phi'_1\rangle=\langle \phi'_2\rangle\neq\langle \phi'_3\rangle\neq0$ then $T'$ is broken into $Z_2$.

\item[(6)] The sixth alignment: $\langle \phi'_1\rangle=\langle
\phi'_2\rangle =\langle \phi'_3\rangle \neq 0$ then $T'$ is broken into $Z_4$.

\item[(7)]  The seventh alignment: $0\neq \langle \phi'_1\rangle\neq \langle
\phi'_2\rangle=\langle \phi'_3\rangle=0$ then $T'$ is broken into $Z_6$.
\end{itemize}

\end{document}